\documentclass{PoS}
\pdfoutput=1

\usepackage{slashed}
\usepackage{amsmath}
\usepackage{graphicx}

\newcommand{\Eq}[1]{Eq.\,\protect\eqref{#1}}
\newcommand{\Fig}[1]{Fig.\,\protect\ref{#1}}
\newcommand{\Tab}[1]{Table\,\protect\ref{#1}}

\title{Quark-photon vertex from lattice QCD in Landau gauge}

\ShortTitle{Quark-photon vertex from lattice QCD in Landau gauge}

\author{Milena Leutnant\\
        Theoretisch-Physikalisches Institut, 
        Friedrich-Schiller-Universit\"at Jena, 07743 Jena, Germany}

\author{\speaker{Andr\'e Sternbeck}\\\\
        Theoretisch-Physikalisches Institut, 
        Friedrich-Schiller-Universit\"at Jena, 07743 Jena, Germany\\
        E-mail: \email{andre.sternbeck@uni-jena.de}}

\abstract{%
We study the nonperturbative structure of the quark-photon vertex in Landau gauge. To this end, we utilize lattice QCD data for the vector current for two mass-degenerate quark flavours and extract all longitudinal and transverse form factors of the underlying vertex for two off-shell kinematics. The momentum dependence of the form factors is compared to the solution of the inhomogeneous Bethe-Salpeter equation for the vertex in the rainbow-ladder approximation. Differences but also similarities are seen between our lattice and the truncated continuum results.}

\FullConference{XIII Quark Confinement and the Hadron Spectrum - Confinement2018\\
		31 July - 6 August 2018\\
		Maynooth University, Ireland}

\begin{document}

\begin{floatingfigure} 
  \includegraphics[width=4.5cm]{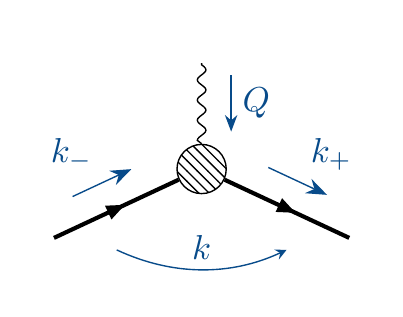}\vspace{-3.5ex}
  \caption{Quark-photon vertex. \label{fig:qpv}}
  \end{floatingfigure}  
\section{Motivation}

Determinations of electromagnetic properties of hadrons using the bound-state approach to QCD require the full, off-shell tensor structure of the quark-photon vertex as input. In particular the case where a virtual photon couples to a strongly interacting off-shell quark and antiquark pair inside a hadron is of interest (see, e.g., \cite{Maris:1999ta,Maris:1999bh,Chang:2013nia,Raya:2015gva,Sanchis-Alepuz:2015fcg,Segovia:2016zyc,Eichmann:2017wil,Weil:2017knt,Raya:2016yuj,Chen:2018rwz,Eichmann:2016yit}). 

For on-shell quarks two form factors are sufficient to parametrize the quark-photon vertex. For an off-shell kinematic there are more degrees of freedom. A common parametrization of the off-shell quark-photon vertex in the continuum reads \cite{Eichmann:2016yit}
\begin{equation}
  \Gamma_\mu(k,Q) = i\gamma_\mu \lambda_1 + 2k^\mu [i\slashed{k}\,\lambda_2 + \lambda_3] \;+\; \sum_{n=1}^8 i f_n T^{(n)}_\mu(k,Q)\;.
  \label{eq:tensorstructure}
\end{equation}
Here $k$ is the average of the respective incoming and outgoing quark momenta, $k_\pm=k\pm Q/2$, and $Q$ is the photon momentum (see \Fig{fig:qpv}). The 11 form factors, $\lambda_{1,2,3}$ and $f_{1,\ldots,8}$, are functions of the Lorentz-invariants $Q^2$, $k^2$ and $\zeta^2=(k\cdot Q)^2/k^2Q^2$, and the base tensors $T^{(n)}_\mu(k,Q)$ are chosen such that the last term in \Eq{eq:tensorstructure} is strictly transverse to $Q$.\footnote{For a definition of the transverse base tensors $T_\mu^{(n)}$ see, e.g., Appendix B of Ref.\,\cite{Eichmann:2016yit}.} This parametrization has the advantage that one obtains $\lambda_{1,2,3}$ directly from the two dressing functions of the nonperturbative quark propagator $S(k_\pm)$ by exploiting the Ward-Takahashi identity (WTI)
\begin{equation}
 Q_\mu \Gamma_\mu(k,Q) = S^{-1}(k_+) - S^{-1}(k_-)\,.
 \label{eq:WTI}
\end{equation}

The transverse form factors, $f_{1,\ldots,8}$, are not fixed by this WTI, but can be found, at least in principle, by solving the inhomogeneous Bethe-Salpeter equation of the quark-photon vertex
\begin{equation}
 \Gamma_\mu(k,Q) = Z_2\gamma_\mu - Z_2^2\frac{4}{3} \int \frac{dq^4}{(2\pi)^4} \left[S(q\!+\!Q/2)\:\Gamma_\mu(q,Q)\:S(q\!-\!Q/2)\right] K(k-q) \,.
 \label{eq:inhBSE}
\end{equation}
Here $Z_2$ denotes the quark wave function renormalisation factor of the quark propagator and $K$ is the quark-antiquark scattering kernel. Solving this equation, however, requires knowledge of the nonperturbative momentum dependences of $S$ and $K$. They satisfy their own Dyson-Schwinger equations (DSEs) and hence involve further $n$-point functions. Strictly speaking, \Eq{eq:inhBSE} leads to an infinite system of equations which for a numerical treatment has to be truncated in a suitable way. 

A frequently used truncation for the quark-photon vertex is the \emph{rainbow-ladder} truncation scheme \cite{Maris:1999ta,Maris:1999bh}. In this scheme the scattering kernel is given by an effective one-gluon exchange, 
\begin{equation}
 K_{\rho\sigma,\alpha\beta}(k) = \gamma_\mu^{\alpha\rho}\, T^k_{\mu\nu}G(k^2) \,\gamma_\nu^{\sigma\beta}
 \qquad\text{with}\quad T^k_{\mu\nu} = \delta_{\mu\nu} - k_\mu k_\nu/k^2\,,
\end{equation}
where $T^k_{\mu\nu}G(k^2)$ is an effective (modelled) gluon propagator and $S(k_\pm)$ is found from a correspondingly truncated quark DSE. In rainbow-ladder truncation one can solve \Eq{eq:inhBSE} numerically (see, e.g., \cite{Maris:1999ta,Maris:1999bh,Eichmann:2014qva}), but the systematic error is hard to control without additional input, for instance from lattice QCD.

\section{Lattice setup}

Lattice QCD provides direct access to the vector-current 3-point function
\begin{equation}     
  G_\mu(k,Q) = \sum_{x,y,z} e^{ik_-(x-z)}    e^{ik_+(z-y)}
  \left\langle D_U^{-1}(x,z) \,i\gamma_\mu\, D_U^{-1}(z,y)    \right\rangle_U
  \qquad  (k_\pm\equiv k\pm Q/2) \label{eq:Gmu}  
\end{equation}  
in Landau gauge.
Here $D^{-1}_U$ denotes the inverse Wilson clover fermion matrix and the gauge
links $U$ satisfy the Landau gauge condition. With corresponding lattice Monte 
Carlo estimates for the quark propagator $S(k_\pm)$ one obtains the vertex
\begin{equation}  
  \Gamma_\mu(k,Q) = S^{-1}(k-Q/2)\; G_\mu(k,Q)\; S^{-1}(k+Q/2) \,.
  \label{eq:GammamuLatt} 
\end{equation} 
for a discrete mesh of momenta, $k$ and $Q$, without having to solve its inhomogeneous BSE. The calculation is similar to a lattice calculation of RI'SMOM renormalisation constants \cite{Sturm:2009kb}, but with the difference that one extracts the full tensor structure of the vertex, that is the dependence of the 11 form factors $\lambda_{1,2,3}$ and $f_{1,\ldots,8}$ on $k^2$, $Q^2$ and $\zeta^2$. These are obtained from a projection of the tensor structure [\Eq{eq:tensorstructure}] onto the lattice result for $\Gamma_\mu$ [\Eq{eq:GammamuLatt}]. The RI'SMOM renormalisation program for this vertex would only target $\lambda_1$ at a fixed renormalisation scale, e.g., $k_\pm^2=Q^2=-\mu^2$.

\begin{table}
 \centering\small
\begin{tabular}{l@{\quad}c@{\quad}c@{\quad}c@{\qquad}c@{\quad}c@{\quad}c@{\quad}r@{\qquad}}
\hline\hline
 no. &$\beta$ & $\kappa$  & $V$ & $a$ [fm] & $Z_V$ & $m_\pi$ [MeV] & \multicolumn{1}{c}{$m$ [MeV]} \\ \hline
C-I & 5.20 & 0.13550 & $32^3\!\times64$ & 0.081 & 0.7219 & $681$ & $36.7$  \\
C-III &  & 0.13584 & $32^3\!\times64$ &  && $409$ & $14.2$  \\
C-IV &  & 0.13596 & $32^3\!\times64$ &  && $280$ & $6.3$  \\*[0.7ex]
E-III & 5.29 & 0.13620 & $32^3\!\times64$ & 0.071 & 0.7364 & $422$ & $17.0$ \\
E-IVs &  & 0.13632  & $32^3\!\times64$ & && $295$ & $8.1$  \\
E-IV  &  & 0.13632  & $64^3\!\times64$ & && $290$ & $8.1$  \\
E-IIV &  & 0.13640   & $64^3\!\times64$ &  && $150$ & $2.1$  \\*[0.7ex]
F-II & 5.40 & 0.13640  & $32^3\!\times64$ & 0.060 & 0.7506 & $490$ & $24.6$ \\
F-III &  & 0.13647  & $32^3\!\times64$ &  && $426$ & $18.4$ \\
F-IV &  & 0.13660  & $48^3\!\times64$ &  && $260$ & $7.0$ \\
\hline \hline
\end{tabular}
 \caption{Our $N_f=2$ gauge ensembles and their parameters. The $m_\pi$-column quotes the respective pion mass and the last column the bare quark mass $m=m_0-m_c(\beta)$. Both $m_\pi$ and $m_c(\beta)$ were provided by the RQCD collaboration \cite{Bali:2016lvx}. The $Z_V$ values are updates of those in \cite{Gockeler:2010yr} and correspond to $r_0=0.5\,\text{fm}$ and $r_0\Lambda^{\overline{\mathsf{MS}}}=0.789$.\label{tab:parameters}}
\end{table}

\begin{figure*}
 \centering
 \mbox{\includegraphics[width=7.4cm]{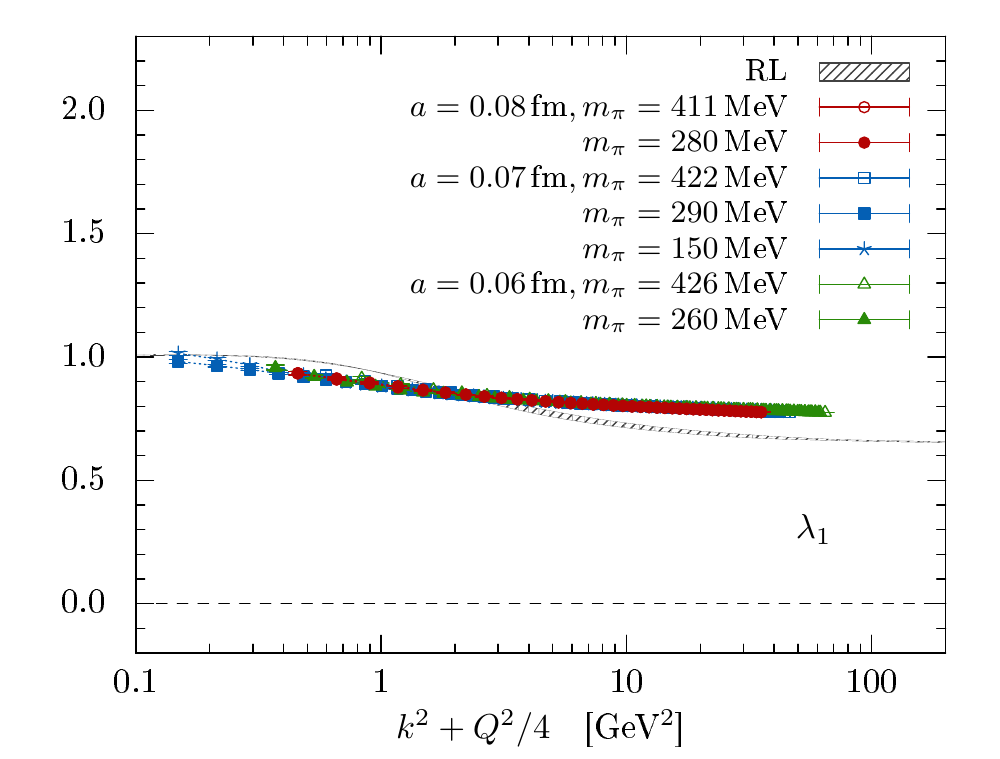}
       \includegraphics[width=7.4cm]{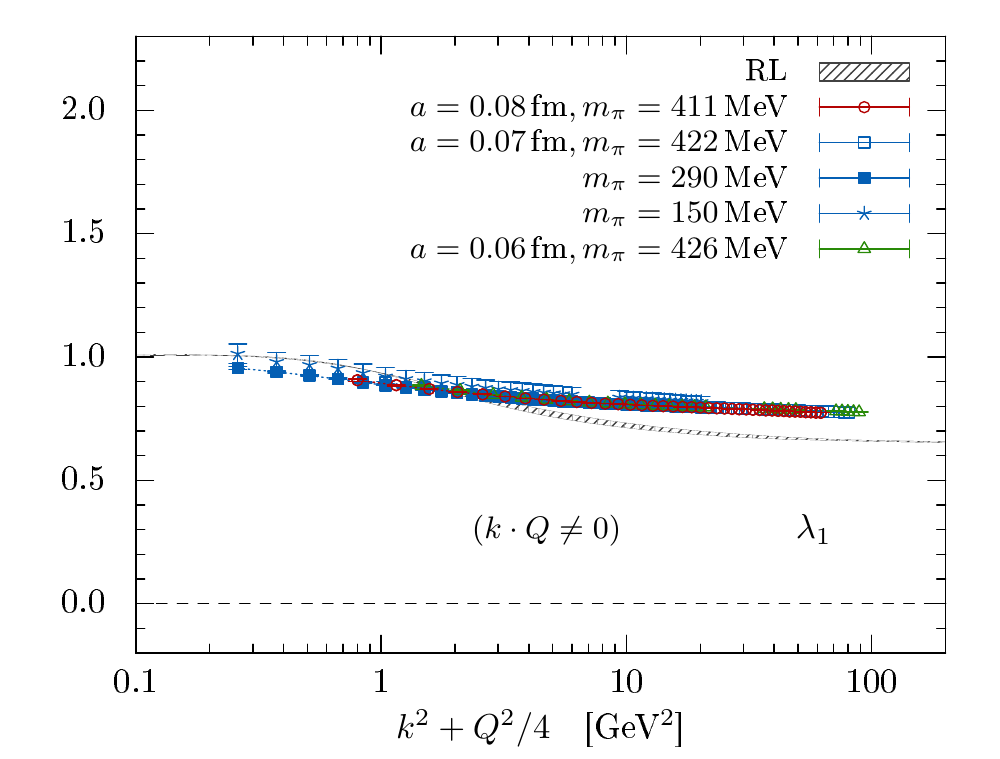}}
 \mbox{\includegraphics[width=7.4cm]{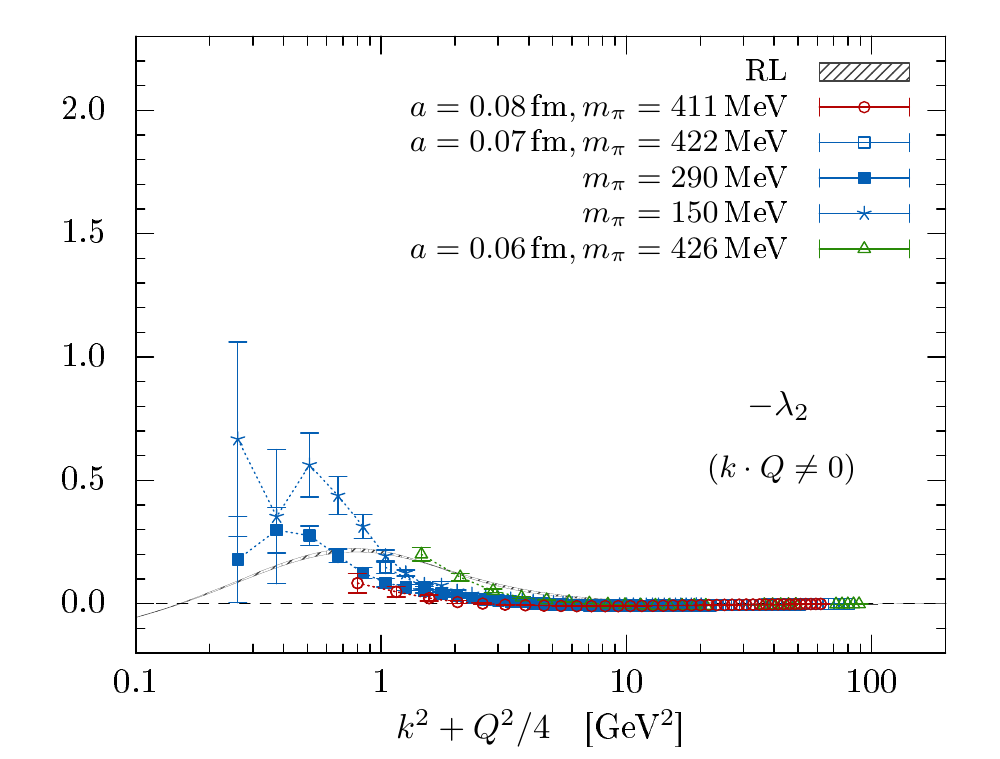}
       \includegraphics[width=7.4cm]{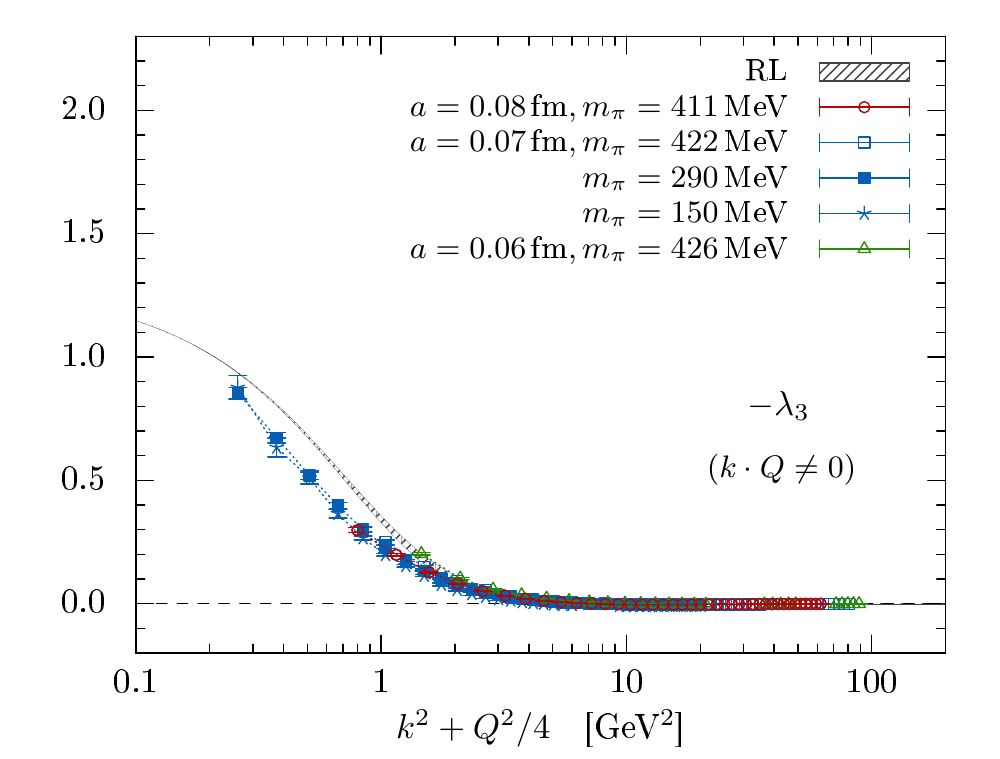}}
\caption{Comparison of our preliminary lattice results (points) to rainbow-ladder (RL) results (bands) for the form factors $\lambda_{1,2,3}$ of the quark-photon vertex from \cite{Eichmann:2014qva,Eichmann:2018priv}. The bands encode the angle dependence of the RL results. For better comparison they have been rescaled by a global factor such that $\lambda_1=1$ for $k^2+Q^2/4\to0$. \label{fig:qpv_long}}
\end{figure*}

\begin{figure*}
 \centering
\mbox{\includegraphics[width=7.2cm]{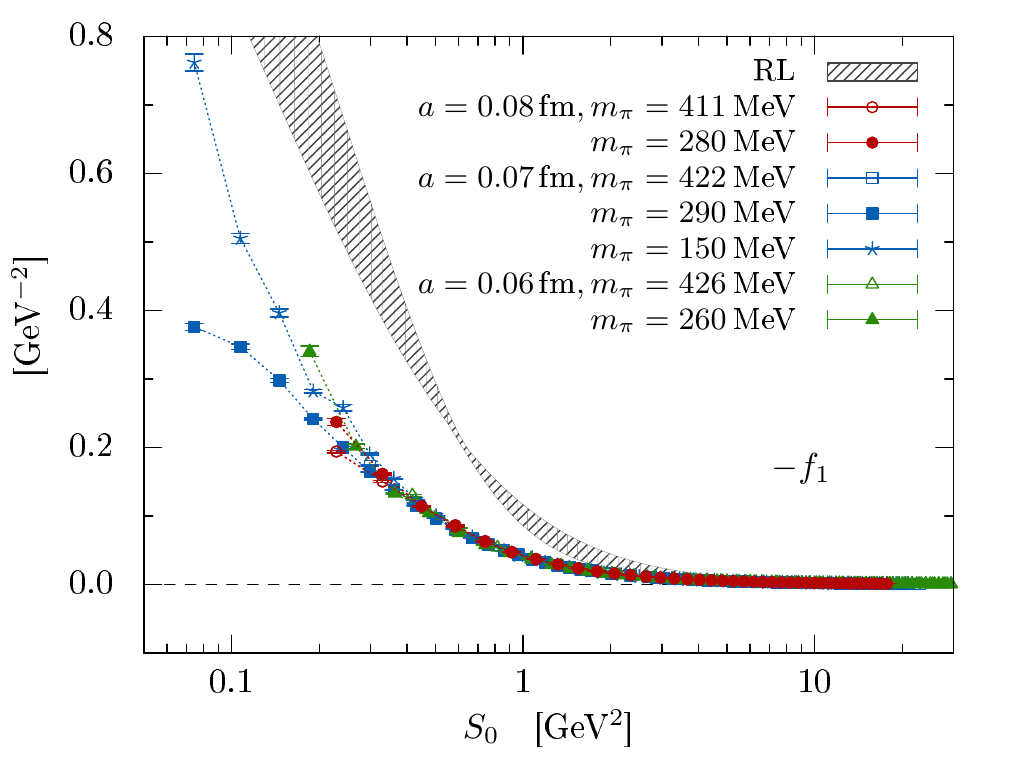}
       \includegraphics[width=7.2cm]{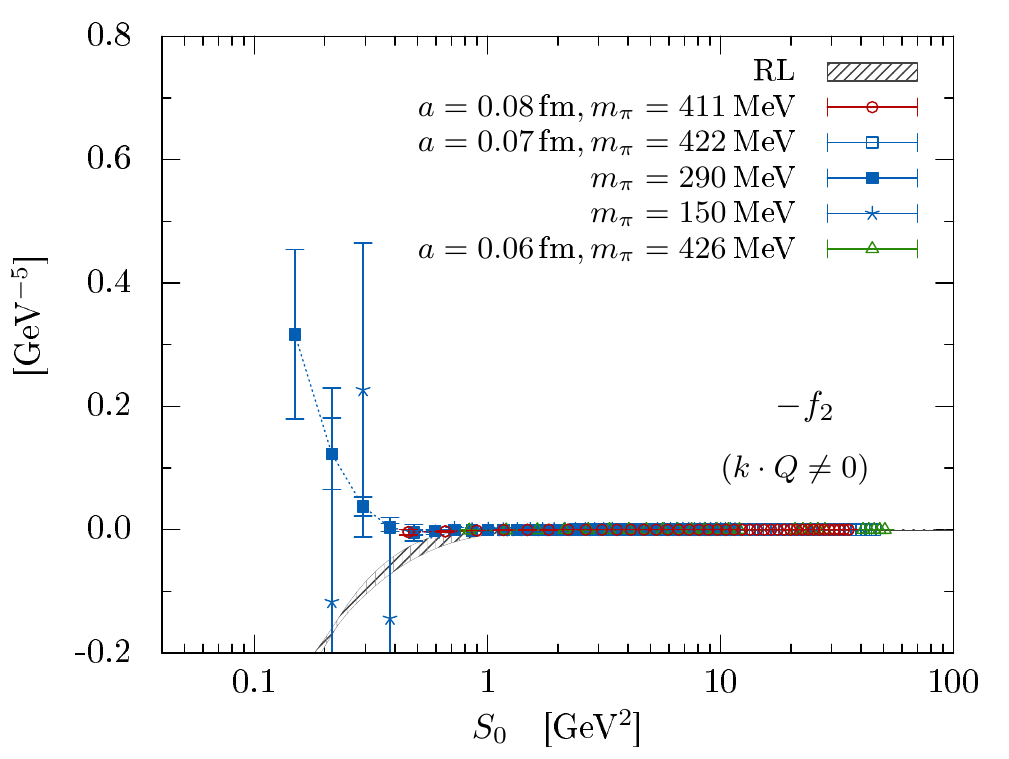}}
 \mbox{\includegraphics[width=7.2cm]{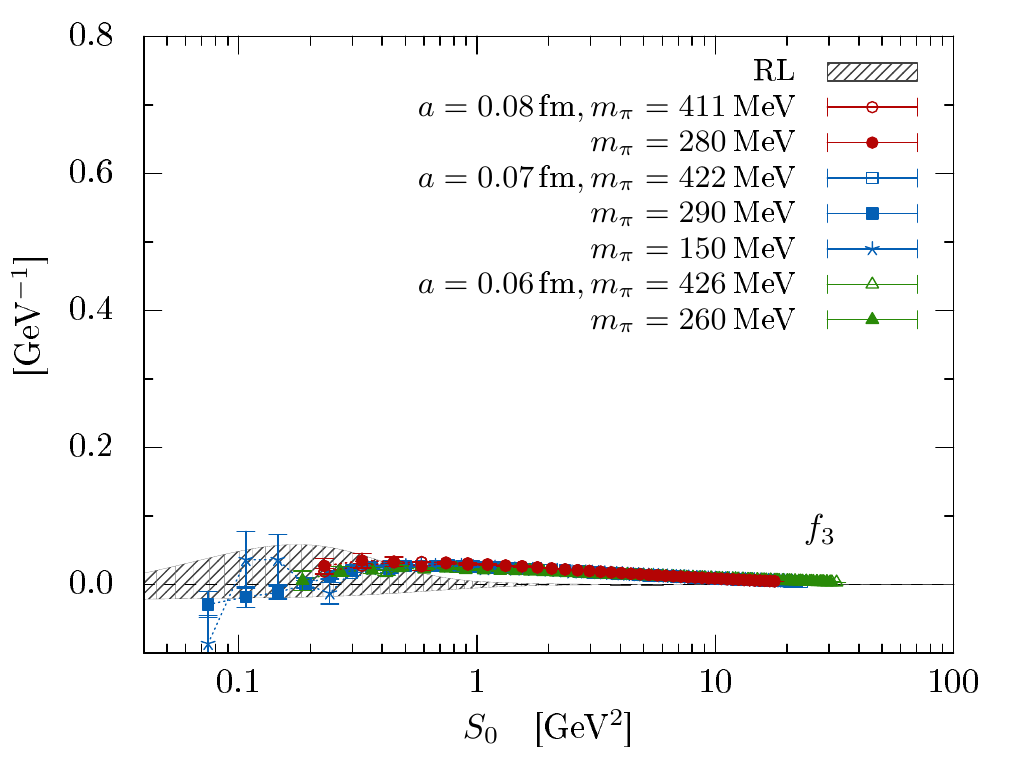}
       \includegraphics[width=7.2cm]{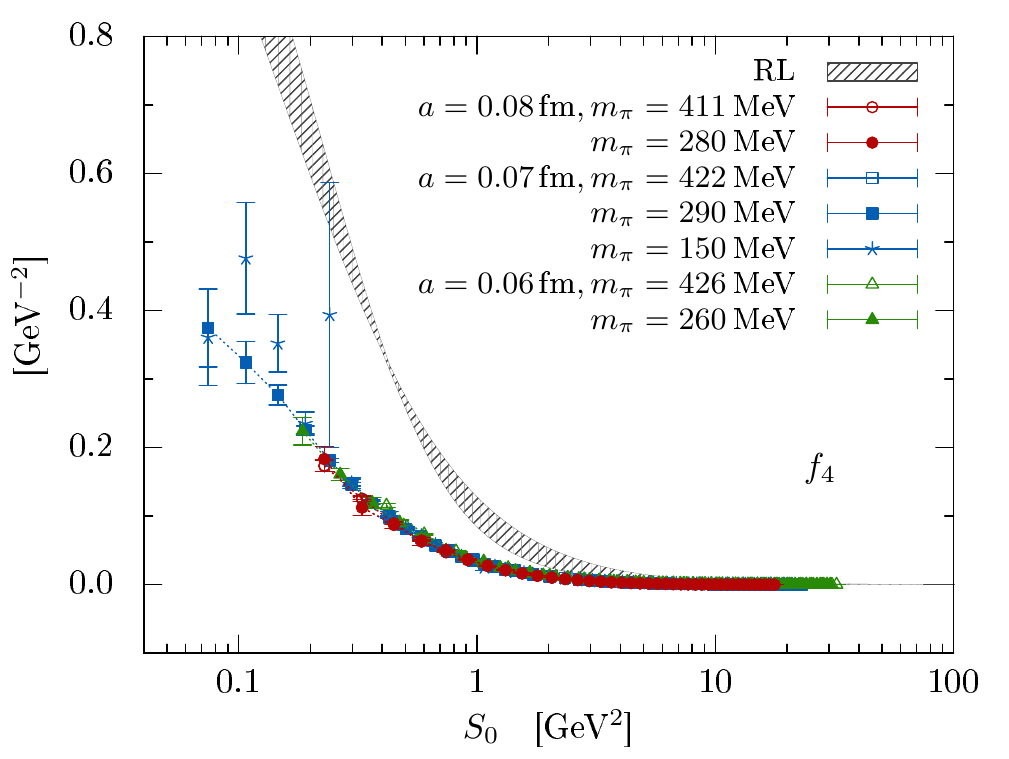}}
 \mbox{\includegraphics[width=7.2cm]{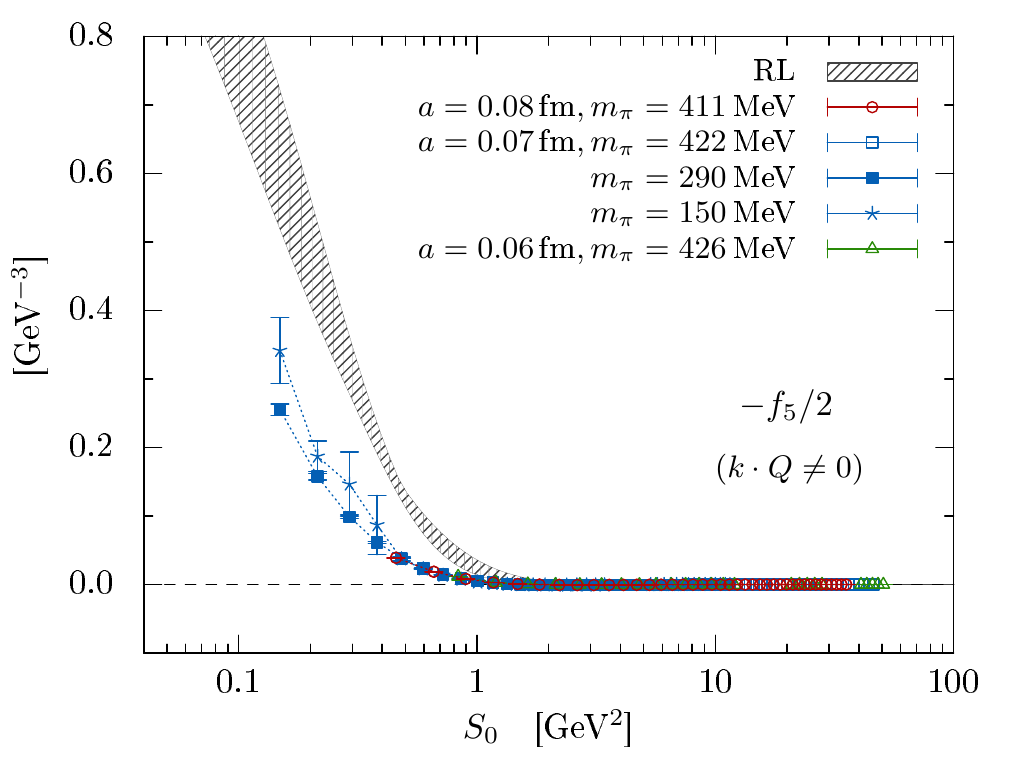}
       \includegraphics[width=7.2cm]{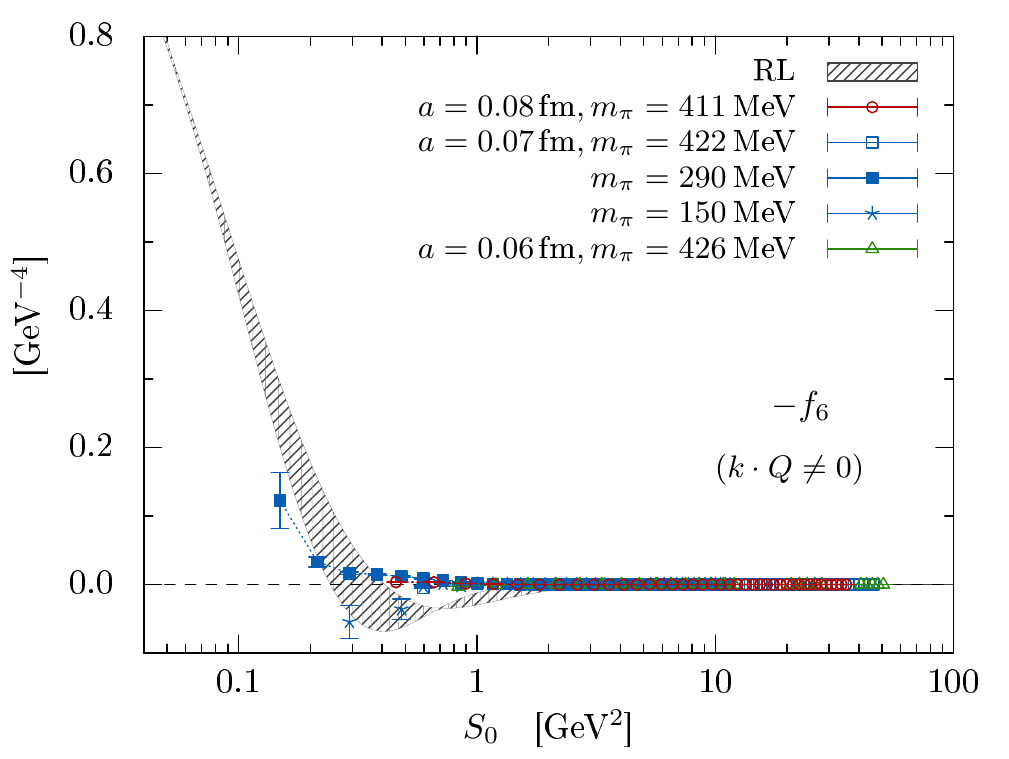}}
 \mbox{\includegraphics[width=7.2cm]{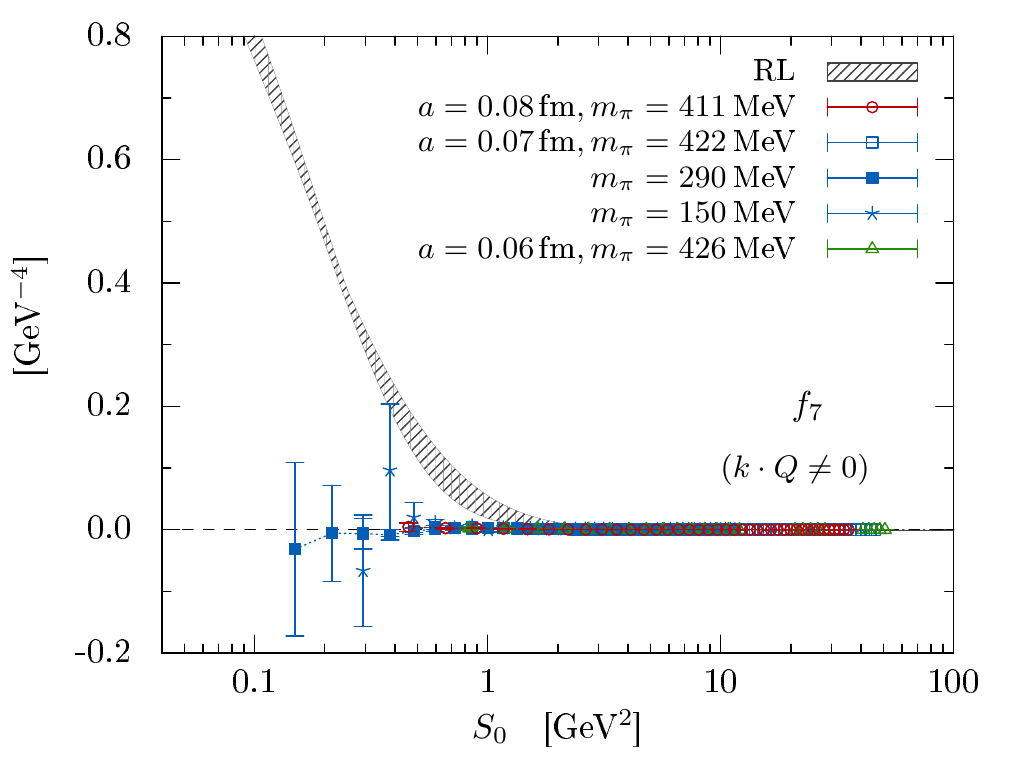}
       \includegraphics[width=7.2cm]{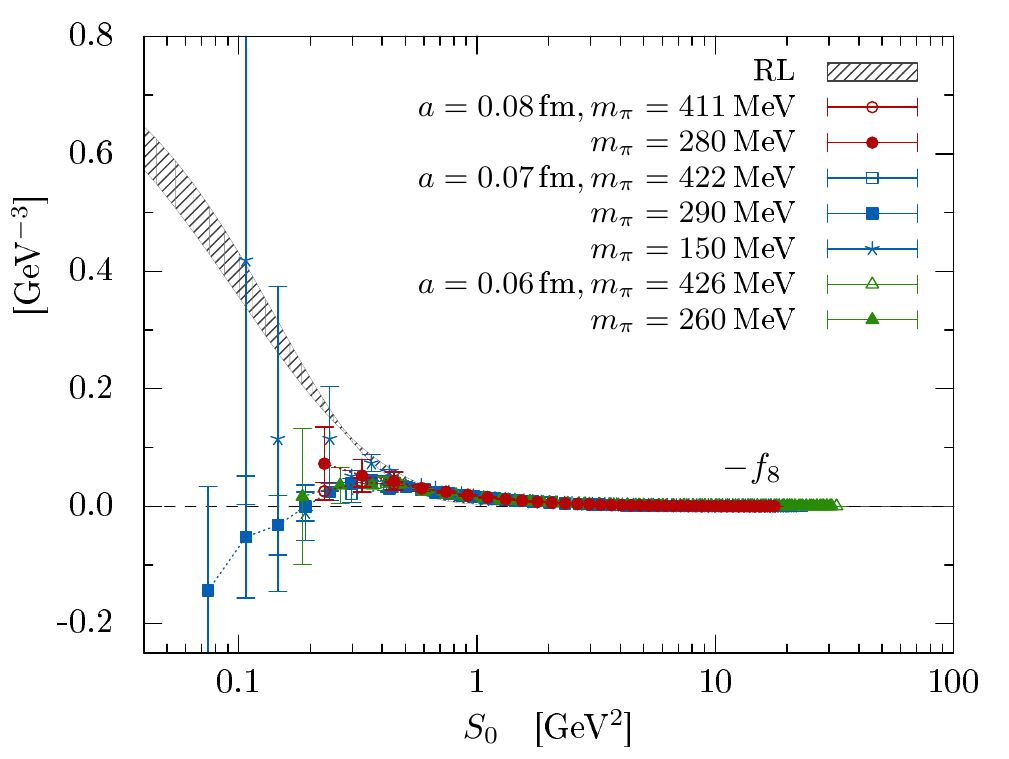}}
 \caption{As \Fig{fig:qpv_long}
  but for the transverse  form factors 
   $f_{1,\ldots,8}$. The lattice results are yet preliminary. \label{fig:qpv_trans}}
\end{figure*}

For the calculation of $G_\mu(k,Q)$ and $S(k_\pm)$ we use momentum-volume 
sources for the inversion of $D_U$. This keeps statistical errors low, even on a few gauge configurations, but requires a separate inversion for each momentum $k_\pm$. We choose two momentum setups: (i) a symmetric setup where $k_-^2=k_+^2=Q^2$ and hence $\zeta = 0$, and (ii) an asymmetric setup where $k_-^2=Q^2 > k_+^2$ and $\zeta = 1/\sqrt{5}$. In addition we employ twisted boundary conditions for the fermions, $\psi(x+\hat{\mu} L_\mu) = e^{i\pi \tau^\mu}\psi(x)$, to refine our momentum mesh. Specifically, we use the following two sets of lattice momenta $ak^\mu_\pm= 2\pi n_\pm^\mu / L_\mu$:
\begin{align}
  \text{(i)}\qquad n_- &= n\,(1,1,0,0) + (\tau,\tau,0,0) & n_+ &= n\,(0,1,1,0) + (0,\tau,\tau,0) \\
  \text{(ii)}\qquad n_- &= n\,(2,1,0,0) + (2\tau,\tau,0,0) & n_+ &= n\,(0,1,1,0)  + (0,\tau,\tau,0) 
\end{align}
with $n=1,2,\ldots,L_s/4$ and $\tau=0,0.4,0.8,1.2$ and $1.6$. Here $L_s$ is the spatial lattice extent which ranges between $L_s=32$ and 64; the temporal extent is always $L_t=64$. Our calculations are performed for three different lattice spacings ($a\simeq 0.06\ldots0.08\,\text{fm}$) and a range of quark masses down to an almost physical mass. This allows us to analyse discretisation, quark mass and volume effects. \Tab{tab:parameters} gives an overview about the parameters of our gauge ensembles.

For our calculation we use the local vector current which violates the WTI [\Eq{eq:WTI}] on the lattice. Hence there is an additional renormalisation factor, $Z_V < 1$, for the vector current, which however is known for our $\beta$ values from other studies. We apply the nonperturbative values from the RQCD collaboration listed in \Tab{tab:parameters}.

\section{Lattice results for the form factors}

Our preliminary lattice results for the form factors $\lambda_{1,2,3}$ and $f_{1,\ldots,8}$ are shown in Figs.\,\ref{fig:qpv_long} and~\ref{fig:qpv_trans}, respectively. For $\lambda_{1,2,3}$ we show them as a function of $k^2+Q^2/4$ and for $f_{1,\ldots,8}$ versus $S_0\equiv k^2/3 + Q^2/4$. This simplifies the comparison with the available continuum results. For $\lambda_{1}$ we show our lattice results for both momentum setups, symmetric and asymmetric. For $\lambda_{2}$ and $\lambda_{3}$ only results for the asymmetric setup are shown. Similar holds for $f_2$, $f_5$, $f_6$ and $f_7$ in \Fig{fig:qpv_trans}. For $f_1$, $f_3$, $f_4$ and $f_8$ we show results for the symmetric setup, but the asymmetric results look alike. Overall we find that the angle ($\zeta^2$) dependence is small. For comparison, both figures also show the corresponding continuum results which were obtained from solving \Eq{eq:inhBSE} in rainbow-ladder (RL) truncation by Eichmann and others \cite{Eichmann:2014qva,Eichmann:2018priv,Sanchis-Alepuz:2017jjd}.  We see qualitative similarities but also quantitative differences, in particular for low momenta.
The largest deviation we see for $f_2$, $f_4$, $f_7$ and $f_8$. For $f_2$ this could be due to a wrong sign in the RL solution, because the RL solution shown in \cite{Sanchis-Alepuz:2017jjd} has the opposite sign which much better fits to the lattice data. Compared to the deviations we see between the RL and lattice results, discretisation or quark mass effects in our lattice data are negligible. Our lattice results therefore provide a very useful guide for the search of improved truncation schemes beyond RL. Still the statistics of our lightest quark-mass ensemble has to be increased to improve the precision at small momentum.

\section{WTI and the quark-photon vertex on the lattice}

Lattice discretisation effects are small compared to the deviations we see between RL and lattice results. However, when looking at the lattice data at a finer resolution, dependencies on both the quark mass $m$ and the lattice spacing $a$ are seen. Let us demonstrate this for the case of $\lambda_1$. This form factor is particularly interesting since in the continuum limit it is solely given by the dressing functions $A(p^2)$ of the quark propagator
\begin{equation}
 S^{-1}(p) = i\slashed{p} A(p^2) + B(p^2) \,,
\end{equation}
while on the lattice---with Wilson fermions and using the local vector current---there are deviations which only disappear with $a$.

\begin{figure*}
  \centering
  \mbox{
    \includegraphics[width=0.5\linewidth]{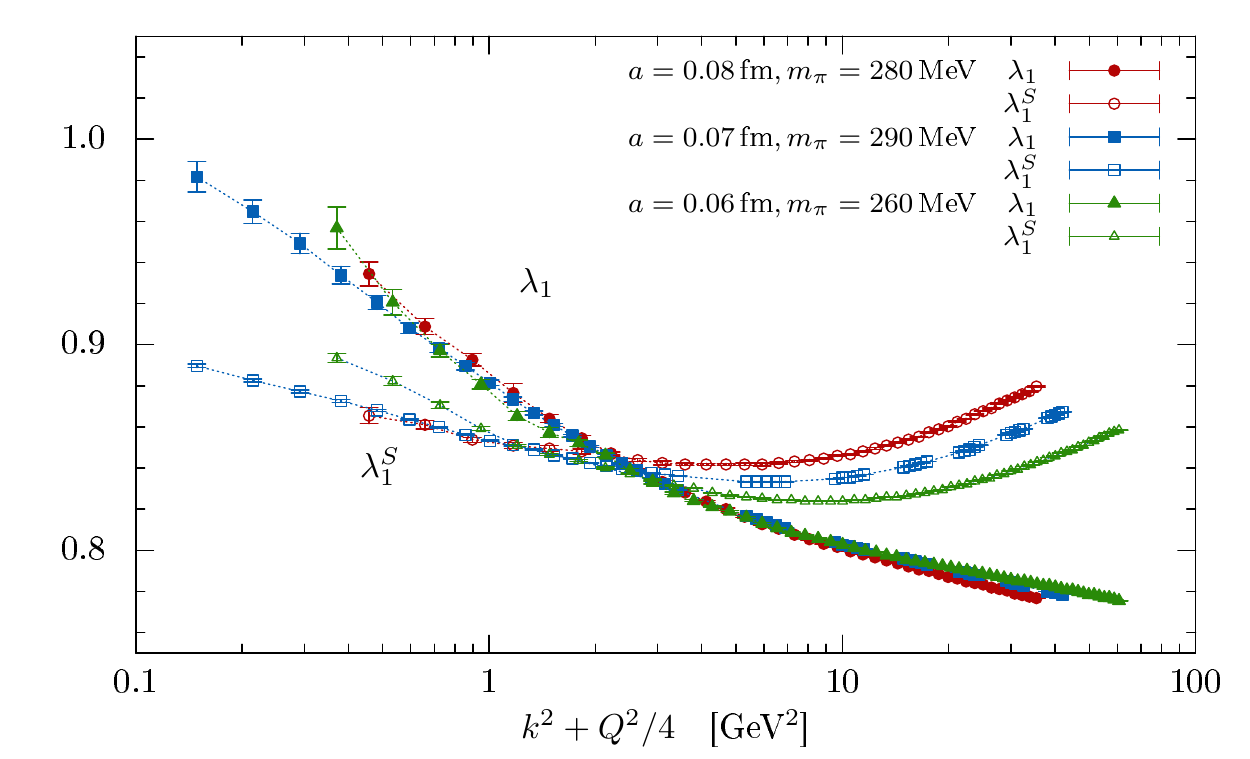}
    \includegraphics[width=0.5\linewidth]{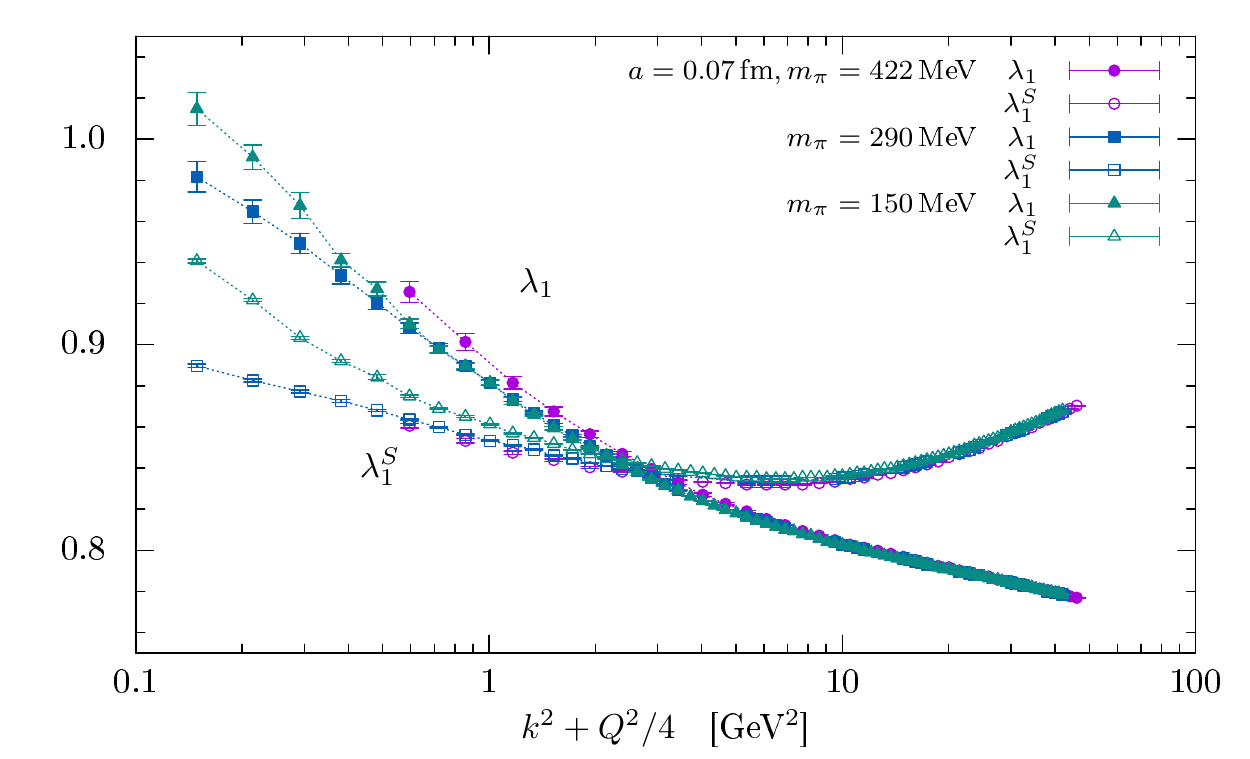}}
  \caption{$\lambda_1$ vs.\ $k^2_\pm$ for symmetric momenta, i.e., $k\cdot Q=0$. The full symbols ($\lambda_1$) are for results extracted from $\Gamma_\mu$, while results with open symbols ($\lambda^S_1$) are obtained from the quark propagator. The left panel shows results for  three different lattice spacings but similar quark masses; the right panel for varying quark masses but fixed lattice spacing.\label{fig:wti}}
\end{figure*}

In \Fig{fig:wti} we show lattice results for $\lambda_1$ vs.\ $k^2_\pm$ for symmetric momenta, extracted either from $\Gamma_\mu$ or $S$ (the latter are the open symbols labelled $\lambda^S_1$). We see that multiplying $\Gamma_\mu$ with $Z_V$ brings $\lambda_1$ and $\lambda^S_1$ in the same ballpark of values, but momentum dependent deviations for small and large momenta remain. Overall, $\lambda_1$ changes more strongly with momentum than $\lambda_1^S$. This discrepancy is expected because the momentum dependent deviation in the vector WTI for the local vector current is not accounted for in $Z_V$ (see, e.g., \cite{Vladikas:2011bp}). For larger momenta the deviations grow with the lattice spacing and momentum, in particular for $\lambda^S_1$. In contrast, for smaller momenta there is a clear dependence on the bare quark mass $am$: for fixed $k^2_\pm$, the gap between $\lambda_1$ and $\lambda^S_1$ grows with $a$ and~$m$.

\begin{floatingfigure}
  \includegraphics[width=0.5\linewidth]{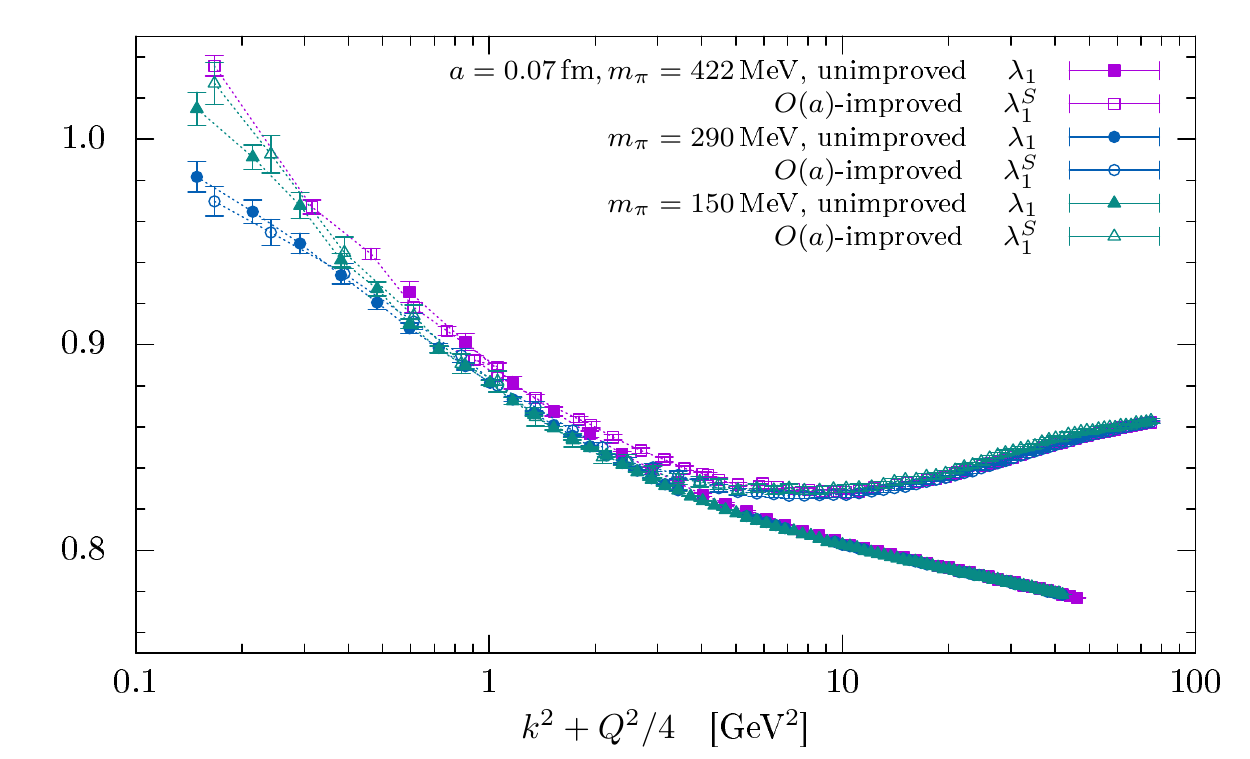}
  \caption{Same as \Fig{fig:wti}, right, but for $\lambda^S_1$ from an $O(a)$-improved Wilson fermion propagator \cite{Oliveira:2018lln}. \label{fig:wti_Oa}}
\end{floatingfigure}
From an independent lattice study \cite{Oliveira:2018lln} we know the quark dressing function of an off-shell $O(a)$-improved Wilson fermion propagator for the same lattice parameters. The corresponding data points for $\lambda_1^S$ are compared to $\lambda_1$ in \Fig{fig:wti_Oa}. We see that the gap at small $k_\pm^2$ is much smaller for the $O(a)$ improved results. An improvement is also seen for larger $k^2_\pm$ but there the effect is comparably smaller.  Unfortunately, we have no data yet for the vector current using the improved quark propagator in \Eq{eq:Gmu}. Nonetheless, it seems a large part of the $O(a)$ effects is cancelled in $\Gamma_\mu$ by the amputation of the external quark legs [\Eq{eq:GammamuLatt}].

\section{Conclusion and outlook}

We have calculated for the first time the nonperturbative QCD corrections to all 11 non-zero form factors of the quark-photon vertex for two off-shell kinematics. Until now only results from a solution of a RL-truncated inhomogeneous Bethe-Salpeter equations were known. Our lattice results give now a first impression for the full solution and hence on the systematic error of the RL truncation. Still our results are afflicted by systematic errors as well, in particular for larger momenta where discretisation artefacts cause deviations of the momentum dependence. Also for small momenta the finite bare quark mass has an effect. These artefacts are however negligible compared to the deviations we see between the RL results and our lattice QCD results. Our lattice results provide therefore a useful guide for the search of improved truncations and on the systematic error of current truncations used for hadron physics calculations within the bound-state approach.

\begin{acknowledgments}
 \noindent We thank Gunnar Bali, Christian Fischer, Meinulf G\"ockeler, Richard Williams and especially Gernot Eichmann and the late Martin Schaden for helpful discussions. We are grateful to the RQCD collaboration for giving us access to their gauge configurations. The gauge fixing and calculations of the fermion propagators were performed on the HLRN supercomputing facilities in Berlin and Hannover (bep00046), the Leibniz Supercomputing Center of the Bavarian Academy of Sciences and Humanities (LRZ) on the supercomputer SuperMUC and at the Ara cluster at the FSU Jena. AS acknowledges support by the BMBF under grant No.\ 05P15SJFAA (FAIR-APPA-SPARC) and by the DFG Research Training Group GRK1523.
\end{acknowledgments}

\bibliographystyle{JHEP}
\bibliography{bibliography}

\end{document}